\documentstyle[12pt]{article}

\begin{document}

\begin{flushright}
IMSc/99/08/** \\
hep-th/9908061
\end{flushright} 

\vspace{2mm}

\vspace{2ex}

\begin{center}
{\large \bf  

             A Note on Holographic Principle in \\ 

\vspace{2ex}
             models of Extended Inflation type } \\ 

\vspace{8ex}

{\large  S. Kalyana Rama}

\vspace{3ex}

Institute of Mathematical Sciences, C. I. T. Campus, 

Taramani, CHENNAI 600 113, India. 

\vspace{1ex}

email: krama@imsc.ernet.in \\ 
\end{center}

\vspace{4ex}

\begin{quote}
ABSTRACT.  

We present a simple derivation of an upper bound on the average
size of the true vacuum bubbles at the end of inflation, in
models of extended inflation type. The derivation uses the
inequality that the total energy inside a given volume must be
less than its linear dimensions. The above bound is the same as
that obtained earlier, by applying the holographic principle
according to Fischler-Susskind prescription. Such a bound leads
to a lower bound on the denisty fluctuations.

\end{quote}

\vspace{2ex}


\newpage

\vspace{4ex}

Holographic principle says, among other things, 
that the entropy in a given 
volume is bounded by its surface area, measured in planck 
units. It is based on the inequality 
\begin{equation}\label{e<r}
E < r 
\end{equation}
where $E$ is the total 
energy inside a given volume, and $r$ is its linear 
dimensions. Otherwise the system would lie within the 
Schwarzschild radius, and would collapse to form 
a black hole \cite{thooft}. 

Recently, Fischler and Susskind (FS) have a given 
a prescription
for applying the holographic principle in cosmology \cite{fs}. 
Applying FS prescription during the inflationary stage, we have 
obtained in an earlier work a lower bound on the density 
fluctuations \cite{rs}.\footnote{ FS prescription is said to be 
invalid during reheating, a non adiabatic process \cite{ven}. 
However, in \cite{rs}, the FS prescription is applied 
only (soon) after the reheating, when the entropy generation 
is complete.} To our 
knowledge, this is the first instance where a {\em lower bound} 
on density fluctuations has been obtained theoretically.  
In the explicit cases considered, this bound is remarkably 
close to the observed value.

Later, other prescriptions for applying the holographic
principle in cosmology have been suggested \cite{others}.
However, none of them seems to lead to a lower bound on 
the density fluctuations. Hence, the existence of such 
a theoretical lower bound, 
important though it is, is in doubt. 

In this note, we present a simple alternative derivation 
of this bound, using equation (\ref{e<r}). 
The derivation is applicable 
to models of extended inflation type, where the inflation 
is ended by the percolation of true vacuum bubbles (and 
eventual reheating) \cite{ei}. 

In the following, we use planck units unless mentioned 
otherwise and, for simplicity, omit numerical factors of 
order unity. Let $T_b$ be 
the temperature when the inflation begins, and 
$\rho_F \simeq T_b^4$ be the false vacuum energy. 
In extended inflation type models, 
the true vacuum bubbles nucleate during inflation, 
expand with the speed of light, and eventually percolate 
the universe completely, thus ending the inflation. 
(The bubble walls then collide and reheat the universe 
to a temperature $T_R \simeq T_b$.) 

Let $d$ be the average size of 
the bubbles at the instant of complete percolation and 
let $\delta \simeq T_b^{- 1}$ be the 
bubble wall thickness. (Thicker bubble walls will 
result in stronger bounds.) 
Then, the energy $E$ stored in the bubble is 
\begin{equation}\label{energy}
E \simeq \pi d^2 \delta \; \rho_F 
\simeq d^2 T_b^3 \; . 
\end{equation} 

Before percolation is complete, the bubbles  
are expanding with the speed of light, and are embedded 
in an inflating 
universe. It seems unlikely that, under these circumstances, 
the bubbles can collapse 
back to form black holes. The inequality (\ref{e<r}) 
may not then be applied. However, upon the completion of 
percolation, the inflation ends, the bubbles stop 
expanding, and are now embedded in an expanding universe 
like ours. Then, regions with sufficient amount of energy 
inside can collapse to form black holes. The inequality 
(\ref{e<r}) may then be applied.

Applying (\ref{e<r}) now gives the constraint 
\begin{equation}\label{hol}
E < d 
\; \; \; \; \longleftrightarrow \; \; \; \; 
d T_b^3 < 1 \; . 
\end{equation}
This constraint is same as that in 
\cite{rs}, which was obtained 
by applying holgraphic principle according to 
Fischler-Susskind prescription. 
As explained in \cite{rs} with explicit examples, 
an upper bound on $d$ leads 
to (i) an upper bound on the duration of inflation; 
(ii) an upper bound on the inflation factor; and 
(iii) a lower bound on the density fluctuations, 
$\frac{\delta \rho}{\rho}$. 

Explicitly, in the case of extended inflation with 
$T_b \simeq 10^{14} GeV$ and $\omega \simeq 10$, one gets  
$\frac{\delta \rho}{\rho} \stackrel{>}{_\sim}
{\cal O} (10^{- 7})$ 
(see \cite{ei, kst, rs} for details). 
This lower bound, obtained theoretically, 
is remarkably close to the observed value 
$\simeq 10^{- 6}$ \cite{cobe}. 

We conclude with two remarks. First, 
it is clear from the above derivation that the violation 
of the bound (\ref{hol}) is likely to 
result in a copius production of black 
holes. We are assuming implicitly that such is not the case. 
On the other hand, it is perhaps of interest to study the  
consequences, cosmological or otherwise, of such a copious 
black hole production. However, such a study lies 
beyond the scope of the present work. 

Second, 
it is not clear if, and how, the above derivation 
can be extended to models of new inflation type, where the 
inflation is ended by the inflaton `rolling down the hill' 
\cite{ni}.



\begin{thebibliography}{999}
\bibitem{thooft} 
G. 't Hooft, gr-qc/9310026; 
L. Susskind, 
J. Math. Phys. {\bf 36} (1995) 6377, hep-th/9409089 and the
references therein. 
\bibitem{fs}
W. Fischler and L. Susskind, hep-th/9806039. 
\bibitem{rs}
S. Kalyana Rama and Tapobrata Sarkar, 
Phys. Lett. {\bf B 450} (1999) 55, hep-th/9812043. 
\bibitem{ven}
G. Veneziano, hep-th/9907012. 
\bibitem{others}
Some of the prescriptions are: 
R. Easther and D. A. Lowe, 
Phys. Rev. Lett. {\bf 82} (1999) 4967, hep-th/9902088; 
G. Veneziano, 
Phys. Lett. {\bf B 454} (1999) 22, hep-th/9902126; 
D. Bak and S.-J. Rey, hep-th/9902173; 
N. Kaloper and A. Linde, hep-th/9904120. 
\bibitem{ei}
D. La and P. J. Steinhardt, 
Phys. Rev. lett. {\bf 62} (1989) 376. 
\bibitem{kst}
E. W. Kolb, D. S. Salopek, and M. S. Turner, 
Phys. Rev. {\bf D 42} (1990) 3925; 
A. H. Guth and B. Jain, 
Phys. Rev. {\bf D 45} (1992) 426. 
\bibitem{cobe}
G. F. Smoot et al, Ap. J. {\bf 396} (1992) L1.  
\bibitem{ni}
A. D. Linde, Phys. Lett. {\em B 108} (1982) 389; 
A. Albrecht and P. J. Steinhardt, 
Phys. Rev. lett. {\bf 48} (1982) 1220. 

\end{thebibliography}
\end{document}